\documentclass{ws-ijmpa}
\usepackage[super,compress]{cite}
\usepackage{graphicx}
\usepackage{url}
\usepackage{hyperref}
\usepackage[utf8]{inputenc}
\begin{document}

\newcommand{\BE}{\begin{equation}}
\newcommand{\EE}{\end{equation}}
\newcommand{\half}{{\scriptstyle{\frac{1}{2}}}}

\bibliographystyle{ws-ijmpa}

\markboth{M. Consoli, L.Cosmai}{The mass scales of the Higgs field}

\catchline{}{}{}{}{}
%
%%%%%%%%%%%%%%%%%%%%%%%%%%%%%%%%%%%%%%%%%%%%%%%%%%%%%%%%%%%%%%%%%%%%

\title{The mass scales of the Higgs field}

\author{Maurizio Consoli}

\address{INFN - Sezione di Catania,  I-95129 Catania, Italy \\
maurizio.consoli@ct.infn.it}

\author{Leonardo Cosmai}

\address{INFN - Sezione di Bari, I-70126 Bari, Italy\\
leonardo.cosmai@ba.infn.it}

\maketitle

\begin{history}
\received{Day Month Year}
\revised{Day Month Year}
\end{history}

\begin{abstract}
In the first version of the theory, with a classical scalar
potential, the sector inducing SSB was distinct from the Higgs field
interactions induced through its gauge and Yukawa couplings. We have
adopted a similar perspective but, following most recent lattice
simulations, described SSB in $\lambda\Phi^4$ theory as a weak
first-order phase transition. In this case, the resulting effective
potential has two mass scales: i) a lower mass $m_h$, defined by its
quadratic shape at the minima, ~and~ ii) a larger mass $M_h$,
defined by the zero-point energy. These refer to different momentum
scales in the propagator and are related by $M^2_h\sim m^2_h \ln
(\Lambda_s/M_h)$, where $\Lambda_s$ is the ultraviolet cutoff of the
scalar sector.  We have checked this two-scale structure with
lattice simulations of the propagator and of the susceptibility in
the 4D Ising limit of the theory. These indicate that, in a cutoff
theory where both $m_h$ and $M_h$ are finite, by increasing the
energy, there could be a transition from a relatively low value,
e.g. $m_h$=125 GeV, to a much larger $M_h$. The same lattice data
give a final estimate $M_h= 720 \pm 30 $ GeV which induces to
re-consider the experimental situation at LHC. In particular an
independent analysis of the ATLAS + CMS data indicating an excess in
the 4-lepton channel as if there were a new scalar resonance around
700 GeV. Finally, the presence of two vastly different mass scales,
requiring an interpolating form for the Higgs field propagator also
in loop corrections, could reduce the discrepancy with those precise
measurements which still favor large values of the Higgs particle
mass.

\end{abstract}

\ccode{PACS numbers: 11.30.Qc; 12.15.-y; 13.85.-t}

%\tableofcontents

\section{Introduction}

Spontaneous Symmetry Breaking (SSB) through the non-vanishing
expectation value $\langle \Phi \rangle \neq$ 0 of a
self-interacting scalar field $\Phi(x)$ is the essential ingredient
to generate the particle masses in the Standard Model. This old idea
\cite{Higgs:1964ia,Englert:1964et} of a fundamental scalar field, in the following
denoted for brevity as the Higgs field, has more recently found an
important experimental confirmation after the observation, at the
Large Hadron Collider of CERN \cite{Aad:2012tfa,Chatrchyan:2012xdj}, of a narrow scalar
resonance, of mass $m_h \sim 125 $ GeV whose phenomenology fits well
with the perturbative predictions of the theory. The discovery of
this resonance, identified as the long sought Higgs boson, has
produced the general conviction that modifications of this general
picture, if any, can only come from new physics, e.g. supersymmetry.

Though, in spite of the present phenomenological consistency, this
conclusion may be too premature. So far only the gauge and Yukawa
couplings of the 125 GeV Higgs particle have been tested. This is
the sector of the theory described by these interactions and by the
associated induced coupling, say $\lambda^{\rm ind}$, determined by
\BE \label{nonauto}\frac{d\lambda^{\rm ind}}{dt}=
\frac{1}{16\pi^2}\left[-12 y^4_t +\frac{3}{4}(g')^4 +
\frac{3}{2}(g')^2g^2 + \frac{9}{4}g^4\right] \EE where $g$ and $g'$
are the SU(2)xU(1) gauge couplings and we have just restricted to
the quark-top Yukawa coupling $y_t$ evolving according to \BE
\label{top}\frac{dy_t}{dt}= \frac{1}{16\pi^2}\left[\frac{9}{2}y_t^3
- \left( \frac{17}{12}(g')^2 + \frac{9}{4}g^2  +8
g^2_3\right)y_t\right] \EE where $g_3$ is the SU(3)$_c$ coupling.
Instead, the effects of a genuine scalar self-coupling $\lambda$, if
any, are below the accuracy of the measurements. For this reason, an
uncertainty about the mechanisms at the base of symmetry breaking
still persists.

We briefly mention that, at the beginning, SSB  was explained in
terms of a classical scalar potential with a double-well shape. Only
later, after the work of Coleman and Weinberg \cite{Coleman:1973jx},
it became increasingly clear that the phenomenon should be described
at the quantum level and that the classical potential had to be
replaced by the effective potential $V_{\rm eff}(\varphi )$ which
includes the zero-point energy of all particles in the spectrum.
This has produced the present view where the description of SSB is
demanded to the combined study of all couplings and of their
evolution up to very large energy scales.

But, in principle, SSB could still be determined by the pure scalar
sector if the contribution of the other fields to the vacuum energy
is negligible. This may happen if, as in the original picture with
the classical potential, the primary mechanism producing SSB is
quite distinct from the remaining Higgs field self-interactions
induced through the gauge and Yukawa couplings. The type of scenario
we have in mind is sketched below:

~~ i) One could first take into account the indications of most
recent lattice simulations of pure $\lambda\Phi^4$ in 4D
\cite{lundow2009critical,Lundow:2010en,akiyama2019phase}.
These calculations, performed in the Ising
limit of the theory with different algorithms, indicate that on the
largest lattices available so far the SSB phase transition is
(weakly) first order.

~~ ii) With a first-order transition, SSB would emerge as a true
instability of the symmetric vacuum at $\varphi=0$. Its quanta have
a tiny and still positive mass squared $V''_{\rm
eff}(\varphi=0)=m^2_\Phi>0$ but, nevertheless, their interactions
can destabilize this symmetric vacuum \cite{Consoli:1999ni}
and produce the condensation process responsible for symmetry
breaking. This primary $\lambda\Phi^4$ sector should be considered
with its own degree of locality defined by some cutoff scale
$\Lambda_s$. We are thus lead to identify $\Lambda_s$ as the Landau
pole for a bare coupling $\lambda_B=+\infty$. This corresponds
precisely to the Ising limit and provides the best possible
definition of a local $\lambda\Phi^4$ for any non-zero low-energy
coupling $\lambda\sim 1/ \ln \Lambda_s\ll 1$. This is the relevant
one for low-energy physics, as in the original Coleman-Weinberg
calculation of the effective potential at $\varphi^2 \ll
\Lambda^2_s$.

~~ iii) After this first step, the description of the basic
$\lambda\Phi^4$ sector can further be improved by going to a next
level. Since, for any non-zero $\lambda$, there is a finite Landau
pole, one can consider the whole set of theories
($\Lambda_s$,$\lambda$), ($\Lambda'_s$,$\lambda'$),
($\Lambda''_s$,$\lambda''$)...with larger and larger Landau poles,
smaller and smaller low-energy couplings but all having the same
depth of the potential, i.e. with the same vacuum energy ${\cal
E}=V_{\rm eff}(\langle \Phi \rangle)$. This requirement derives from
imposing the RG-invariance of the effective potential in the
three-dimensional space ($\varphi$, $\lambda$, $\Lambda_s$) and, in
principle, allows one to handle the $\Lambda_s \to \infty$ limit
\footnote{This limit should also be considered because the scalar
sector is assumed to induce SSB and thus to determine the vacuum
structure and its symmetries. In a quantum field theory, imposing
invariance under RG-transformations is then the standard method to
remove the ultraviolet cutoff or, in alternative, to minimize its
influence on observable quantities.}. In this formalism, besides a
first invariant mass scale ${ \cal I}_1$, defined by $|{\cal E}|\sim
{ \cal I}^4_1$, there is a second invariant ${\rm \cal I}_2$,
related to a particular normalization of the vacuum field, which is
the natural candidate to represent the weak scale ${ \cal
I}_2=\langle \Phi \rangle\sim$ 246 GeV. The minimization of the
effective potential can then be expressed as a relation ${ \cal
I}_1= K { \cal I}_2$ in terms of some proportionality constant $K$.

This RG-analysis of the effective potential, discussed in Sects.2
and 3, is the main point of this paper. It takes into account that,
in those approximation schemes that reproduce the type of weak
first-order phase transition favored by recent lattice simulations,
there are {\it two} vastly different mass scales, say $m_h$ and
$M_h$. These are defined respectively by the second derivative and
the depth of the effective potential at its minima and related by
$M^2_h\sim L m^2_h >> m^2_h$ where $L=\ln (\Lambda_s/M_h)$.
Therefore, even though $(m_h/\langle \Phi \rangle)^2 \sim 1/L$, the
larger $M_h={ \cal I}_1$ remains finite in units of ${ \cal
I}_2=\langle \Phi \rangle$.

To appreciate the change of perspective, let us recall the usual
description of a second-order phase transition as summarized in the
scalar potential reported in the Review of Particle Properties
\cite{Tanabashi:2018oca}. In this review, which gives the present
interpretation of the theory in the light of most recent
experimental results, the scalar potential is expressed as
(PDG=Particle Data Group) \BE \label{VPDG} V_{\rm
PDG}(\varphi)=-\frac{ 1}{2} m^2_{\rm PDG} \varphi^2 + \frac{
1}{4}\lambda_{\rm PDG}\varphi^4 \EE By fixing $m_{\rm PDG}\sim$ 88.8
GeV and $\lambda_{\rm PDG}\sim 0.13$, this potential has a minimum
at $|\varphi|=\langle \Phi \rangle\sim$ 246 GeV and quadratic shape
$V''_{\rm PDG}(\langle \Phi \rangle)=$ (125 GeV)$^2$. Note that, as
a built-in relation, the second derivative of the potential (125
GeV)$^2$ also determines its depth, i.e. the vacuum energy ${\cal
E}_{\rm PDG}$
 \BE \label{EPDG} {\cal E}_{\rm PDG}=-\frac{ 1}{2} m^2_{\rm PDG}
\langle \Phi\rangle^2 + \frac{ 1}{4}\lambda_{\rm PDG} \langle
\Phi\rangle^4 =-\frac{ 1}{8} (125~{\rm GeV}\langle \Phi\rangle)^2
\sim -1.2\cdot10^8~ {\rm GeV}^4\EE Instead in our case, by
identifying $m_h\sim$ 125 GeV, the vacuum energy ${\cal E}\sim
-\frac { 1}{8} M^2_h \langle \Phi \rangle^2$ would be deeper than
Eq.(\ref{EPDG}) by the potentially divergent factor $L$. Thus, it
would also be insensitive to the other sectors of the theory, e.g.
the gauge and Yukawa interactions, whose effect is just to replace
the scalar self coupling $\lambda$ with the total coupling
$\lambda^{\rm tot}=\lambda +\lambda^{\rm ind}$ in the definition of
the quadratic shape of the effective potential. All together, once
the picture sketched above works also in the $\Lambda_s \to \infty$
limit, where $\lambda$ becomes extremely small at any finite energy
scale, the phenomenology of the 125 GeV resonance would remain the
same and SSB would essentially be determined by the pure scalar
sector.

We emphasize that the relation $M_h= K\langle \Phi \rangle$ is not
introducing a new large coupling $K^2=O(1)$ in the picture of
symmetry breaking. This $K^2$  should not be viewed as a coupling
constant or, at least, as a coupling constant which produces {\it
observable} interactions in the broken symmetry phase. From this
point of view, it may be useful to compare SSB to the phenomenon of
superconductivity in non-relativistic solid state physics. There the
transition to the new, superconductive phase represents an essential
instability that occurs for any infinitesimal two-body attraction
$\epsilon$ between the two electrons forming a Cooper pair. At the
same time, however, the energy density of the superconductive phase
and all global quantities of the system (energy gap, critical
temperature, etc.) depend on the much larger collective coupling
$\epsilon N$ obtained after re-scaling the tiny 2-body strength by
the large number of states near the Fermi surface. This means that,
in principle, the same macroscopic description could be obtained
with smaller and smaller $\epsilon$ and Fermi systems of
corresponding larger and larger $N$. In this comparison $\lambda$ is
the analog of $\epsilon$ and $K^2$ is the analog of $\epsilon N$.

Another aspect, implicit in the usual picture of SSB, is that
$V''_{\rm PDG}( \langle \Phi \rangle )$, which strictly speaking is
the self-energy function at zero momentum $|\Pi(p=0)|$, is assumed
to coincide with the pole of the Higgs propagator. As discussed in
Sect.4, $m_h$ and $M_h$ refer to different momentum regions in the
connected scalar propagator $G(p)=1/ (p^2 -\Pi(p))$, namely $m_h$
for $p \to 0$ and $M_h$ at larger $p$. Therefore, if $\Lambda_s$
were large but finite, so that both $m_h$ and $M_h$ are finite, the
transition between the two scales should become visible by
increasing the energy.

In Sect.5, we will show that this two-scale structure is supported
by lattice simulations in the 4D Ising limit of the theory. In fact,
once $m^2_h$ is directly computed from the zero-momentum connected
propagator $G(p=0)$ (the inverse susceptibility) and $M_h$ is
extracted from the behaviour of $G(p)$ at higher momentum, the
lattice data confirm the increasing expected logarithmic trend
$M^2_h\sim L m^2_h$.

From a phenomenological point of view, these simulations indicate
that a relatively low value, e.g. $m_h$=125 GeV, could in principle
coexist with a much larger $M_h$. By combining various lattice
determinations, our final estimate $M_h= 720 \pm 30 $ GeV will lead
us to re-consider, in Sect.6, the experimental situation at LHC. In
particular, an independent analysis \cite{Cea:2018tmm} of the ATLAS
+ CMS data indicating an excess in the 4-lepton channel as if there
were a new scalar resonance around 700 GeV. This excess, if
confirmed, could indicate the second heavier mass scale discussed in
this paper. Then, differently from the low-mass state at 125 GeV,
the decay width of such heavy state into longitudinal vector bosons
will be crucial to determine the strength of the {\it observable}
scalar self-coupling and the degree of locality of the theory.

Finally, the simultaneous presence of two mass scales would also
require an interpolating parametrization for the Higgs field
propagator in loop corrections. This could help to reduce the
3-sigma discrepancies with those precision measurements which still
favor rather large values of the Higgs particle mass.

\section{The one-loop effective potential}

To study SSB in $\lambda\Phi^4$ theory, the crucial quantity is the
physical, mass squared parameter $m^2_\Phi=V''_{\rm eff}(\varphi=0)$
introduced by first quantizing the theory in the symmetric phase at
$\varphi=0$. A first-order scenario corresponds to a phase
transition occurring at some small but still positive $m^2_\Phi$. In
this case, the symmetric vacuum, although {\it locally} stable
(because its excitations have a physical mass $m^2_\Phi> 0$), would
be {\it globally} unstable in some range of mass below a critical
value, say $0 \leq m^2_\Phi <m^2_c$. If  $m^2_c$ is extremely small,
however, one speaks of a {\it weak} first-order transition to mean
that it would become indistinguishable from a second-order
transition if one does not look on a fine enough scale.

This first-order scenario is equivalent to say that the lowest
energy state of the massless theory at $m^2_\Phi=0$ corresponds to
the broken-symmetry phase, as suggested by Coleman and Weinberg
\cite{Coleman:1973jx} in their one-loop calculation. This represents the
simplest scheme which is consistent with this picture. We will first
reproduce below this well known computation and exploit its
implications. A discussion on the general validity of the one-loop
approximation is postponed to the following section.

The Coleman-Weinberg potential is \BE V_{\rm eff}(\varphi) =
\frac{\lambda}{4!} \varphi^4 +\frac{\lambda^2}{256 \pi^2} \varphi^4
\left[ \ln (\half \lambda \varphi^2 /\Lambda^2_s ) - \frac{1}{2}
\right]  \EE and its first few derivatives are \BE \label{vprime}
%\label{final}
V'_{\rm eff}(\varphi) =  \frac{\lambda}{6} \varphi^3
+\frac{\lambda^2}{64 \pi^2} \varphi^3 \ln (\half \lambda \varphi^2
/\Lambda^2_s )  \EE and \BE \label{vsecond}
%\label{final}
V''_{\rm eff}(\varphi) =  \frac{\lambda}{2} \varphi^2
+\frac{3\lambda^2}{64 \pi^2} \varphi^2 \ln (\half \lambda \varphi^2
/\Lambda^2_s ) +\frac{\lambda^2\varphi^2}{32\pi^2}   \EE We observe
that, by introducing the mass squared parameter \BE
M^2(\varphi)\equiv \half \lambda\varphi^2 \EE the one-loop potential
can be expressed as a classical background + zero-point energy of a
particle with mass $M(\varphi)$ (after subtraction of constant terms
and of quadratic divergences), i.e. \BE \label{zero} V_{\rm
eff}(\varphi) = \frac{\lambda\varphi^4}{4!} - \frac{M^4(\varphi)
}{64 \pi^2}  \ln \frac{ \Lambda^2_s \sqrt{e} } {M^2(\varphi)}   \EE
Thus, non-trivial minima of $V_{\rm eff}(\varphi)$ occur at those
points $\varphi=\pm v$ where \footnote{In view of a possible
ambiguity in the normalization of the vacuum field, that may affect
the identification of the weak scale $\langle \Phi\rangle\sim$ 246
GeV, we will for the moment denote as $\varphi=\pm v$ the minima
entering the computation of the effective potential.}\BE
\label{basic} M^2_h \equiv M^2(\pm v)={{\lambda v^2}\over{2
}}=\Lambda^2_s \exp( -{{32 \pi^2 }\over{3\lambda }})\EE so that \BE
\label{mh} m^2_h\equiv V''_{\rm eff}(\pm v) =
\frac{\lambda^2v^2}{32\pi^2}= \frac{\lambda}{16\pi^2}M^2_h\sim
\frac{M^2_h}{L} \ll M^2_h \EE where $L\equiv \ln
\frac{\Lambda_s}{M_h}$.  Notice that the energy density depends on
$M_h$ and {\it not} on $m_h$, because \BE \label{basicground} {\cal
E}= V_{\rm eff}(\pm v)= -\frac{M^4_h}{128 \pi^2 } \EE therefore the
critical temperature at which symmetry is restored, $k_BT_c\sim
M_h$, and the stability conditions of the broken phase depends on
the larger $M_h$ and not on the smaller scale $m_h$.

These are the results for the $m_\Phi=0$ case. To study the phase
transition for a small $m^2_\Phi >0$, we will just quote the results
of Ref.\cite{Consoli:1999ni}. In this case, the one-loop potential
has the form \BE \label{nonzero}V_{\rm eff}(\varphi)= \half
m^2_\Phi\varphi^2+ \frac{\lambda\varphi^4}{4!}+ \frac{M^4(\varphi)
}{64 \pi^2} \left[ \ln \frac{M^2(\varphi) } { \sqrt{e} \Lambda^2_s}
+ F\left(\frac{m^2_\Phi} { M^2(\varphi)}\right) \right]  \EE where
\BE F(y)= \ln (1+y) + \frac{y(4+3y)} { 2 (1+y)^2} \EE Then, by
introducing the mass-squared parameter Eq.(\ref{basic}) of the $
m_\Phi=0$ case, the condition for non-trivial minima $\varphi= \pm
v$ for $ m_\Phi \neq 0$, can be expressed as \cite{Consoli:1999ni}
\BE m^2_\Phi \leq \frac{\lambda M^2_h} { 64\pi^2\sqrt{e}} \equiv
m^2_c \EE Since the critical mass for the phase transition vanishes,
in units of $M_h$, in the $\Lambda_s \to \infty$ limit \BE
\frac{m^2_c} { M^2_h }\sim \frac{1} { L} \to 0\EE SSB emerges as an
infinitesimally weak first-order transition.

Notice that this critical mass has the same typical magnitude as the
quadratic shape $m^2_h$ in Eq.(\ref{mh}). In this sense, by
requiring SSB, we are establishing a mass hierarchy
\cite{Consoli:1999ni}. On the one hand, the tiny mass of the
symmetric phase $m^2_\Phi\leq m^2_c$ and the similar infinitesimal
quadratic shape $m^2_h$ of the potential at its minima. On the other
hand, the much larger $M^2_h$ entering the zero-point energy which
destabilizes the symmetric phase \footnote{The analysis for the
one-component scalar field can be easily extended to a continuous
symmetry O(N) theory. To this end, it is convenient to follow
ref.\cite{Dolan:1974gu} where it is shown that the one-loop
potential is only due to the zero-point energy associated with the
radial field $\rho (x)$, the contribution from the Goldstone bosons
being exactly canceled by the change in the quantum measure $(Det
\rho)$.}.

As anticipated in the Introduction, to improve our analysis of the
primary $\lambda\Phi^4$ sector, we will now consider the whole set
of pairs ($\Lambda_s$,$\lambda$),($\Lambda'_s$,$\lambda'$),
($\Lambda''_s$,$\lambda''$)...with different Landau poles and
corresponding low-energy couplings. The correspondence is such to
obtain the same value for the vacuum energy Eq.(\ref{basicground}),
or equivalently for the the mass scale Eq.(\ref{basic}), and thus
the cutoff independence of the result by requiring \BE
\label{CSground} \left(\Lambda_s\frac{\partial}{\partial\Lambda_s} +
\Lambda_s\frac{\partial \lambda}{\partial\Lambda_s}\frac{\partial
}{\partial \lambda}\right){\cal E}(\lambda,\Lambda_s)=0 \EE By
assuming Eq.(\ref{basicground}) and with the definition \BE
\Lambda_s\frac{\partial \lambda}{\partial\Lambda_s}\equiv
-\beta(\lambda)= -\frac{3\lambda^2}{16\pi^2} +O(\lambda^3)  \EE the
solution is thus $|{\cal E}| \sim {\cal I }^4_1$, where ${\cal I
}_1$ is the first RG-invariant \footnote{ Note the minus sign in the
definition of the $\beta-$ function. This is because we are
differentiating the coupling constant
$\lambda=\lambda(\mu,\Lambda_s)$, at a certain scale $\mu=M_h$ and
with cutoff $\Lambda_s$, with respect to the cutoff and not with
respect to $\mu$. Namely, at fixed $\mu$, we are considering
different integral curves so that $\lambda$ has to decrease by
increasing $\Lambda_s$. Also, to use consistently the 1-loop
$\beta-$function in Eq.(\ref{I1}), the integral at the exponent
should be considered a definite integral that only depends on
$\lambda$ because its other limit, say $\lambda_0> \lambda$, is kept
fixed and such that, for $x<\lambda_0$, one can safely neglect
$O(x^3)$ terms in $\beta(x)$. Therefore, since $\lambda_0$ cannot be
too large, there is a relative $\lambda-$independent factor
$\exp({{16 \pi^2 }\over{3\lambda_0 }})>>1$ between Eq.(\ref{basic})
and Eq.(\ref{I1}). Strictly speaking, this means that, to obtain the
same physical $M_h$ from Eq.(\ref{basic}) and Eq.(\ref{I1}), one
should use vastly different values of $\Lambda_s$. This is a typical
example of cutoff artifact.} \BE \label{I1} {\cal I }_1= M_h=
\Lambda_s \exp({\int^{\lambda}\frac{dx}{\beta(x)} })\sim \Lambda_s
\exp( -{{16 \pi^2 }\over{3\lambda }})\EE The above relations derive
from the more general requirement of RG-invariance of the effective
potential in the three-dimensional space ($\varphi$, $\lambda$,
$\Lambda_s$) \BE \label{CSveff}
\left(\Lambda_s\frac{\partial}{\partial\Lambda_s} +
\Lambda_s\frac{\partial \lambda}{\partial\Lambda_s}\frac{\partial
}{\partial \lambda}  + \Lambda_s\frac{\partial
\varphi}{\partial\Lambda_s}\frac{\partial }{\partial \varphi}
\right)  V_{\rm eff}(\varphi,\lambda,\Lambda_s)=0 \EE In fact, at
the minima $\varphi=\pm v$, where $(\partial V_{\rm eff}/\partial
\varphi)= 0$, Eq.(\ref{CSground}) is a direct consequence of
Eq.(\ref{CSveff}).

Another consequence of this RG-analysis is that, by introducing an
anomalous dimension for the vacuum field \BE \label{anomalous}
\Lambda_s\frac{\partial \varphi}{\partial\Lambda_s}\equiv
\gamma(\lambda) \varphi \EE there is a second invariant associated
with the RG-flow in the ($\varphi$, $\lambda$, $\Lambda_s$) space,
namely \BE \label{I2} {\cal I }_2(\varphi) = \varphi
\exp({\int^{\lambda}dx \frac{\gamma(x)}{\beta(x)} })\EE which
introduces a particular normalization of $\varphi$. This had to be
expected because from Eq.(\ref{basic}) the cutoff-independent
combination is \BE \lambda v^2\sim M^2_h={\cal I }^2_1 \EE and not
$v^2$ itself, thus implying $\gamma= \beta/(2\lambda)$ \footnote{We
emphasize that this is the anomalous dimension of the {\it vacuum}
field $\varphi$ which is the argument of the effective potential. As
such, it is quite unrelated to the more conventional anomalous
dimension of the {\it shifted} field as obtained from the residue of
the connected propagator $Z=Z_{\rm prop}= 1 + O(\lambda)$. By
``triviality'', the latter is constrained to approach unity in the
continuum limit. To better understand the difference, it is useful
to regard symmetry breaking as a true condensation phenomenon
\cite{Consoli:1999ni} associated with the macroscopic occupation the
same quantum state ${\bf k}=0$. Then $\varphi$ is related to the
condensate while the shifted field is related to the modes at ${\bf
k} \neq 0$ which are not macroscopically populated. Numerical
evidence for these two different re-scalings will be provided in
Sect.5. In fact, the logarithmic increasing $L$ relating $v^2$ and
$\langle \Phi \rangle^2$ is the counterpart
\cite{Consoli:1993jr,Consoli:1997ra} of the logarithmic increasing
$L$ between $M^2_h$ and $m^2_h$ which can be observed on the
lattice.}. Therefore, the condition for the minimum of the effective
potential can be expressed as a proportionality relation between the
two invariants in terms of some constant $K$, say \BE{\cal I }_1= K
{\cal I }_2(v) \EE Then, with the aim of extending our description
of SSB to the Standard Model, a question naturally arises. Suppose
that, as in the first version of the theory, SSB is essentially
generated in the pure scalar sector and the other couplings are just
small perturbative corrections. When we couple scalar and gauge
fields, and we want to separate the field in a vacuum component and
a fluctuation, which is the correct definition of the weak scale
$\langle \Phi \rangle\sim $ 246 GeV? A first possibility would be to
identify $\langle \Phi \rangle$ with the same $v$ considered so far
which in general, i.e. beyond the Coleman-Weinberg limit, is related
to $M_h$ through a relation similar to Eq.(\ref{basic}), say \BE
\label{large} v^2\sim L M^2_h= L {\cal I }^2_1 \EE But $\langle \Phi
\rangle \sim $ 246 GeV is a basic entry of the theory (as the
electron mass and fine structure constant in QED). For such a
fundamental quantity, once we are trying to describe SSB in a
cutoff-independent way, it would be more appropriate a relation with
the second invariant, i.e. \BE \langle \Phi\rangle^2 = {\cal
I}^2_2(v) = \frac {{\cal I}^2_1}{K^2 }= \frac{ M^2_h}{K^2 } \EE so
that both $\langle \Phi \rangle^2\sim (v^2/L)$ and $M^2_h\sim
(v^2/L)$ are cutoff-independent quantities. If we adopt this latter
choice, the proportionality can then be fixed through the
generalization of Eq.(\ref{mh}) in terms of some constant $c_2$ \BE
\label{basicnew} V''_{\rm eff}(\pm v) = m^2_h \sim \frac{c_2
M^2_h}{L}\EE and the traditional definition of $\langle \Phi\rangle$
from the quadratic shape of the effective potential \BE V''_{\rm
eff}(\pm v) = m^2_h = \frac{ \lambda\langle \Phi\rangle^2}{3}\sim
\frac{16 \pi^2}{9L} \langle \Phi\rangle^2 \EE This gives \BE
\label{kfinite} M_h \sim \frac{4\pi}{3 \sqrt{c_2}} \langle \Phi
\rangle\equiv K \langle \Phi \rangle\EE in terms of the constant
$c_2$ that, in Sect.5, will be estimated from lattice simulations of
the theory.

\section{On the validity of the one-loop potential }

Following the lattice simulations of
refs.\cite{lundow2009critical,Lundow:2010en,akiyama2019phase}, which
support the picture of SSB in $\lambda\Phi^4$ as a weak first-order
transition, we have considered in Sect.2 the simplest approximation
scheme which is consistent with this scenario, namely the one-loop
effective potential. From its functional form and its minimization
conditions, we have also argued that this simplest scheme can become
the basis for an alternative approach to the ideal continuum limit
such that the vacuum energy ${\cal E}$ and the natural definition of
the Standard Model weak scale $\langle \Phi \rangle \sim $ 246 GeV
are both finite, cutoff independent quantities.

But one may object that, as remarked by Coleman and Weinberg already
in 1973, the straightforward minimization procedure followed in our
Sect.2, and used to derive ${\cal E}$ and $\langle \Phi \rangle$,
can be questioned. The point is that by performing the standard
Renormalization Group (RG) ``improvement'' of the one-loop
potential, all leading-logarithmic terms are reabsorbed into a
positive running coupling constant $\lambda(\varphi) $. Thus, by
preserving the positivity of $\lambda(\varphi) $, the one-loop
minimum disappears and one would now predict a second-order
transition at $m^2_\Phi = 0$, as in the classical potential. The
conventional view is that the latter result is trustworthy while the
former is not. The argument is that the one-loop potential's
non-trivial minimum occurs where the one-loop ``correction'' term is
as large as the tree-level term. However, also this standard
RG-improved result can be questioned because, near the one-loop
minimum, the convergence of the resulting geometric series of
leading logs is not so obvious.

To gain insight, one can then compare with other approximation
schemes, for instance the Gaussian approximation
\cite{Barnes:1978cd,Stevenson:1985zy} which has a variational nature
and explores the Hamiltonian in the class of the Gaussian functional
states. It also represents a very natural alternative because, at
least in the continuum limit, a Gaussian structure of Green's
functions fits with the generally accepted ``triviality'' of the
theory in 3+1 dimensions. This other calculation produces a result
in agreement with the one-loop potential
\cite{Consoli:1993jr,Consoli:1997ra}. This agreement does not mean
that there are no non-vanishing corrections beyond the one-loop
level; there are, but those additional terms do not alter the
functional form of the result. The point is that, again, as in the
one-loop approximation, the gaussian effective potential can be
expressed as a classical background + zero-point energy with a
$\varphi-$dependent mass as in Eq.(\ref{zero}) \footnote{As already
remarked for the one-loop potential, also for the Gaussian effective
potential the zero-point energy in a spontaneously broken O(N)
theory is just due to the shifted radial field. For the Gaussian
approximation this requires the diagonalization
\cite{Naus:1995xu,Okopinska:1995su} of the mass matrix to explicitly
display a spectrum with one massive field and (N-1) massless fields
as required by the Goldstone theorem.}, i.e. \BE \label{vgauss}
%\label{final}
V^G_{\rm eff}(\varphi) =  \frac{\hat\lambda\varphi^4}{4!}
-\frac{\Omega^4(\varphi) }{64 \pi^2}  \ln \frac{ \Lambda^2_s
\sqrt{e} } {\Omega^2(\varphi)} \EE with \BE \hat \lambda=
\frac{\lambda } {1 + \frac{\lambda}{16 \pi^2} \ln \frac {\Lambda_s}{
\Omega(\varphi)}   }  \EE and  \BE \Omega^2(\varphi)    =
\frac{\hat\lambda\varphi ^2} {2 } \EE This explains why the one-loop
potential can also admit a non-perturbative interpretation. It is
the prototype of the gaussian and post-gaussian calculations
\cite{Stancu:1989sk,Cea:1996pe} where higher-order contributions to
the energy density are effectively reabsorbed into the same basic
structure: a classical background + zero-point energy with a
$\varphi-$dependent mass.
\begin{figure}
\begin{center}
\includegraphics[bb=20 0 1100 500,
angle=0,scale=0.23]{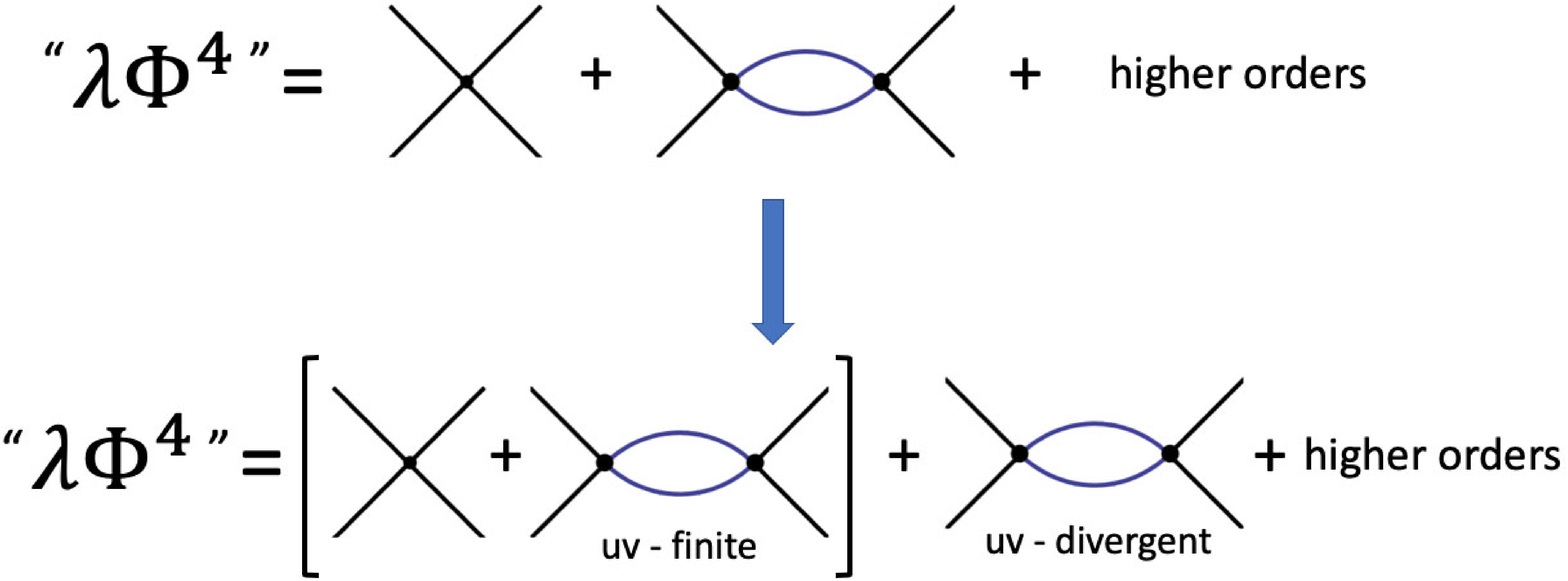} \caption{\it The
re-arrangement of the perturbative expansion considered by Stevenson
\cite{Stevenson:2008mw} in his alternative RG-analysis of the
effective potential. Besides the tree-level +$\lambda \delta^{3}(\bf
r)$ repulsion, the quanta of the symmetric phase, with mass
$m_\Phi$, feel a $-\lambda^2 \frac {e^{-2 m_\Phi r}}{r^3}$
attraction from the Fourier transform of the second diagram in
square bracket \cite{Consoli:1999ni} whose range becomes longer and
longer in the $m_\Phi \to 0$ limit. For $m_\Phi$ below a critical
mass $m_c$, this dominates and induces SSB in the one-loop
potential. Since the higher-order terms just renormalize these two
basic effects, the RG-improved effective potential, in this new
scheme, confirms the same scenario of the one loop approximation. }
\end{center}
\label{pictorial}
\end{figure}
But, even by taking into account the indications of lattice
simulations
\cite{lundow2009critical,Lundow:2010en,akiyama2019phase}, and having
at hand the explicit one-loop and gaussian calculations
Eqs.(\ref{zero}) and (\ref{vgauss}), a skeptical reader may still be
reluctant to abandon the standard second-order scenario. He would
like a general argument explaining why the standard RG-analysis,
which predicts the correct $\Lambda_s-$dependence of the low-energy
coupling, fails instead to predict the order of the phase
transition.

Finding such a general argument was, indeed, the motivation of
ref.\cite{Consoli:1999ni} : understanding the physical mechanisms at
the base of SSB as a first-order transition. Here, the crucial
observation was that the quanta of the symmetric phase, the
``phions'' \cite{Consoli:1999ni}, besides the +$\lambda
\delta^{3}(\bf r)$ tree-level repulsion, also feel a $-\lambda^2
\frac {e^{-2 m_\Phi r}}{r^3}$ attraction which shows up at the
one-loop level and whose range becomes longer and longer in the
$m_\Phi \to 0$ limit \footnote{Starting from the scattering matrix
element $\cal M$, obtained from Feynman diagrams, one can construct
an interparticle potential that is is basically the 3-dimensional
Fourier transform of $\cal M$, see the articles of Feinberg et al.
\cite{Feinberg:1968zz,Feinberg:1989ps}.}. By taking into account
both effects, a calculation of the energy density in the dilute-gas
approximation \cite{Consoli:1999ni}, which is equivalent to the
one-loop potential, indicates that for small $m_\Phi$ the
lowest-energy state is not the empty state with no phions but a
state with a non-zero density of phions Bose condensed in the
zero-momentum mode. The instability corresponds to spontaneous
symmetry breaking and happens when the phion's physical mass
$m^2_\Phi$ is still positive.

Then, if one thinks that SSB originates from these two qualitatively
different competing effects, one can now understand why the standard
RG-resummation fails to predict the order of the phase transition.
In fact, the one-loop attractive term originates from the {\it
ultraviolet finite} part of the one-loop diagrams. Therefore, the
correct way to include higher order terms in the effective potential
is to renormalize {\it both} the tree-level repulsion and the
long-range attraction, as in a theory with {\it two} coupling
constants \footnote{This is similar to what happens in scalar
electrodynamics \cite{Coleman:1973jx}. There, if the scalar self-coupling is not
too large, no conflict arises between one-loop potential and its
standard RG-improvement. }. This strategy, which is clearly
different from the usual one, has been implemented by Stevenson
\cite{Stevenson:2008mw}, see Fig.1. In this new scheme, one can obtain
SSB without violating the positivity of $\lambda(\varphi)$ so that
one-loop effective potential and its RG-group improvement now agree
very well. Stevenson's analysis confirms the weak first-order
scenario and the same two-mass picture $M^2_h\sim m^2_h \ln
(\Lambda_s/M_h)$.

\section{$m_h$ and $M_h$: the quasi-particles of the broken phase}

After having described the various aspects and the general validity
of the one-loop calculation, let us now try to sharpen the meaning
of the two mass scales $m_h$ and $M_h$. To this end, we will first
express the inverse propagator in its general form in terms of the
2-point self-energy function $\Pi(p)$ \BE \label{inverse} G^{-1}(p)=
p^2 -\Pi(p) \EE Then, since the derivatives of the effective
potential produce (minus) the n-point functions at zero external
moment, our smaller mass can be expressed as \BE \label{Pi0}
m^2_h\equiv V''_{\rm eff}(\varphi=\pm v)=-\Pi(p=0)=|\Pi(p=0)| \EE so
that $G^{-1}(p)\sim p^2 + m^2_h $ for $p\to 0$.

As far as $M_h$ is concerned, we can instead use the relation of the
zero-point-energy (``zpe'') in Eq.(\ref{zero}) to the trace of the
logarithm of the inverse propagator \BE\label{general} zpe=
\frac{1}{2} \int {{ d^4 p}\over{(2\pi)^4}} \ln (p^2-\Pi(p))\EE Then,
after subtraction of constant terms and of quadratic divergences, to
match the one-loop form in Eq.(\ref{zero}), we can impose suitable
lower and upper limits to the $p$-integration in the logarithmic
divergent part (i.e. $p^2_{\rm max}\sim \sqrt{e}\Lambda^2_s$ and
$p^2_{\rm min}\sim M^2_h$)
 \BE \label{connection2} zpe=-\frac{1}{4} \int^{p_{\rm
max}}_{p_{\rm min}} {{ d^4 p}\over{(2\pi)^4}} \frac{\Pi^2(p)}{p^4}
\sim-\frac{ \langle \Pi^2(p)\rangle }{64\pi^2} \ln\frac{p^2_{\rm
max}}{p^2_{\rm min}}\sim -\frac{M^4_h}{64\pi^2}
\ln\frac{\sqrt{e}\Lambda^2_s }{M^2_h} \EE This shows that the
quartic term $M^4_h$ is associated with the typical, average value
$\langle \Pi^2(p)\rangle$ at non-zero momentum. Thus, if we trust in
the one-loop relation $M^2_h\sim m^2_h\ln\frac{\Lambda_s}{M_h}$,
there should be substantial deviations when trying to extrapolate
the propagator to the higher-momentum region with the same
1-particle form $G^{-1}(p)\sim p^2 + m^2_h $ which controls the
$p\to 0$ limit.

Before considering deviations of the propagator from the standard
1-particle form, one should first envisage what kind of constraints
are placed by ``triviality''. This dictates a continuum limit as a
generalized free-field theory, i.e. where all interaction effects
are reabsorbed into the first two moments of a Gaussian
distributions. Therefore, in this limit, the spectrum can just
contain free massive particles.

However Stevenson's alternative RG-analysis \cite{Stevenson:2008mw},
besides confirming the two-scale structure $M^2_h\sim m^2_h \ln
(\Lambda_s/M_h)$ found at one loop, also indicates how to recover
the massive free-field limit in an unconventional way. In fact, his
propagator interpolates between $G^{-1}(p=0)= m^2_h $ and
$G^{-1}(p)\sim (p^2 + M^2_h) $ at momenta $p^2 >> m^2_h$, see his
Eqs.(16)$-$(22). This suggests the general following form of the
propagator \BE \label{interpolation} G^{-1}(p) = (p^2 + M^2_h) f (p)
\EE with $f(p)\sim (m_h/M_h)^2$ in the $p \to 0$ limit and $f (p)
\to 1$ for momenta $p^2 >> m^2_h$. Also, note that his Eq.(23)
should be read as $G^{-1}(p)$ and that he considers the continuum
limit $(m_h/M_h)^2 \to 0$. Then $f(p)$ becomes a step function which
is unity for any finite $p$ (i.e. for any $p$ finite in units of
$M_h$) except for a discontinuity at $p=0$ where $f=0$. Up to this
discontinuity in the zero-measure set $p=0$, one then re-discovers
the usual trivial continuum limit with just one massive free
particle \footnote{Note that $p=0$ represents a Lorentz-invariant
set being transformed into itself under any transformation of the
Poincar\'e Group. Thus, in principle, a continuum limit with a
discontinuity in the zero-measure set $p=0$ is not forbidden in
translational invariant vacua as with SSB.}.

We are thus lead to consider the following picture of the cutoff
theory where both $m_h$ and $M_h$ are finite, albeit vastly
different scales. This picture introduces two types of
``quasi-particles'':  quasi-particles of type I, with mass $m_h$,
and quasi-particles of type II, with mass $M_h$. The quasi-particles
of type I are the weakly coupled excitations of the broken-symmetry
phase in the low-momentum region. By increasing the momentum these
first quasi-particle states become more strongly coupled. However,
the constraint placed by ``triviality'' is that, by approaching the
continuum limit, all interaction effects have to be effectively
reabsorbed into the mass of other quasi-particles, those of type II,
i.e. into the parameter we have called $M_h$. The very large
difference between $M_h$ and $m_h$, expected from our analysis of
the effective potential, implies that at higher momentum the
self-coupling of quasi-particles of type I becomes substantial but,
nevertheless, will remain hidden in the transition from $m_h$ to
$M_h$. In an ideal continuum limit, the whole low-momentum region
for the quasi-particles of type I reduces to the zero-measure set
$p=0$ and one is just left \footnote{Here, an analogy can help
intuition. To this end, one can compare the continuum limit of SSB
to the incompressibility limit of a superfluid. In general, this has
two types of excitations: low-momentum compressional modes (phonons)
and higher momentum vortical modes (rotons). If the sound velocity
$c_s\to \infty$ the phase space of the phonon branch, the analog of
the quasi-particles of type I, with energy $E({\bf k})=c_s|{\bf
k}|$, would just reduce to the zero-measure set ${\bf k}=0$. Then,
in this limit, only rotons, the analog of the quasi-particles of
type II, would propagate in the system.} with the quasi-particles of
type II with mass $M_h$.

To show that this new interpretation of ``triviality'' is not just
speculation, in the following section, we will report the results of
lattice simulations of the broken-symmetry phase which support our
two-mass picture.

\section{Comparison with lattice simulations}

We will now compare the two-mass picture of Sects.2-4 with the
results of lattice simulations in the broken-symmetry phase of
$\lambda\Phi^4$ in 4D.
These simulations have been performed in the
Ising limit of the theory governed by the lattice action
\begin{equation}
\label{ising} S_{\rm{Ising}} = -\kappa \sum_x\sum_{\mu} \left[
\phi(x+\hat e_{\mu})\phi(x) + \phi(x-\hat e_{\mu})\phi(x) \right]
\end{equation}
with the lattice field $\phi(x)$ taking only the values $\pm 1$.
Also, the broken-symmetry phase corresponds to $\kappa> \kappa_c$,
this critical value being now precisely determined as
$\kappa_c=0.0748474(3)$ \cite{lundow2009critical,Lundow:2010en}.

Addressing to \cite{Lang:1993sy,montvay1997quantum} for the various
aspects of the analysis, we recall that the Ising limit is
traditionally considered a convenient laboratory for a
non-perturbative study of the theory. As anticipated in the
Introduction, it corresponds to a $\lambda\Phi^4$ with an infinite
bare coupling, as if one were sitting precisely at the Landau pole.
In this sense, for any finite cutoff, it provides the best
definition of the local limit for a given value of the renormalized
parameters.

Using the Swendsen-Wang \cite{Swendsen:1987ce} and
Wolff~\cite{Wolff:1988uh} cluster algorithms, we computed the vacuum
expectation value
\begin{equation}
\label{baremagn}
 v=\langle |\phi| \rangle \quad , \quad \phi \equiv \frac{1}{V_4}\sum_x
\phi(x)
\end{equation}
and the connected propagator
\begin{equation}
\label{connected} G(x)= \langle \phi(x)\phi(0)\rangle - v^2
\end{equation}
where $\langle ...\rangle$ denotes averaging over the lattice
configurations.

\begin{figure}[ht]
\centering
\includegraphics[width=0.8\textwidth,clip]{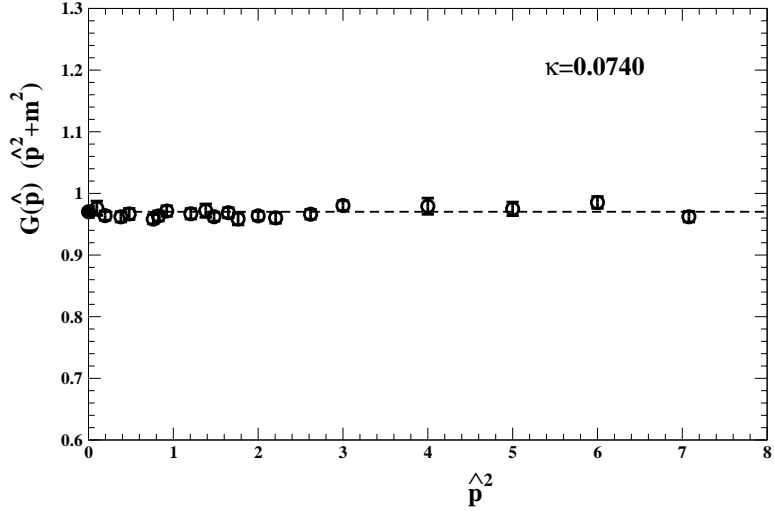}
\caption{\it The lattice data for the re-scaled propagator in the
symmetric phase at $\kappa=0.074$ as a function of the square
lattice momentum ${\hat p}^2$ with $\hat{p}_\mu =2 \sin p_\mu/2$.
The fitted mass is $m_{\rm{latt}}=0.2141(28)$ and the dashed line
indicates the value of $Z_{\rm{prop}}=0.9682(23)$. The zero-momentum
full point is
$Z_\varphi=(2\kappa\chi_{\rm{latt}})m^2_{\rm{latt}}=0.9702(91)$.
Data are taken from Ref.~\cite{Cea:1999kn}. } \label{0.074}
\end{figure}

Our scope was to check the basic relation $M^2_h\sim m^2_h \ln
(\Lambda_s/M_h)$ where $M_h$ describes the higher momentum
propagator and $m_h$ is defined from the zero-momentum 2-point
function Eq.(\ref{Pi0}) \BE m^2_h\equiv V''_{\rm eff}(\pm
v)=-\Pi(p=0)= |\Pi(p=0)|\EE By introducing the Fourier transform of
the propagator $G(p)$, its $p=0$ limit is the susceptibility $\chi$
whose conventional definition includes the normalization factor
$2\kappa$, i.e.  $2\kappa\chi \equiv 2\kappa G(p=0)$. Therefore the
extraction of $m_h$ is straightforward \BE 2\kappa \chi=2\kappa
G(p=0) =\frac{1}{|\Pi(p=0)|} \equiv \frac{1}{m^2_h} \EE Extraction
of $M_h$ requires more efforts. To this end, let us denote by $m
_{\rm{latt}}$ the mass obtained directly from a fit to the
propagator data in some region of momentum. If our picture is
correct, the difference of the value $M_h\equiv m _{\rm{latt}}$, as
fitted in the higher-momentum region, from the corresponding
$m_h\equiv (2\kappa \chi_{\rm{latt}})^{-1/2}$, should become larger
and larger in the continuum limit. Namely, the quantity \BE
Z_\varphi=\frac{M^2_h}{m^2_h} \equiv m^2 _{\rm{latt}} (2\kappa
\chi_{\rm{latt}}) \EE should exhibit a definite logarithmic increase
when approaching the critical point $\kappa \to \kappa_c$.

\begin{figure}[ht]
\centering
\includegraphics[width=0.8\textwidth,clip]{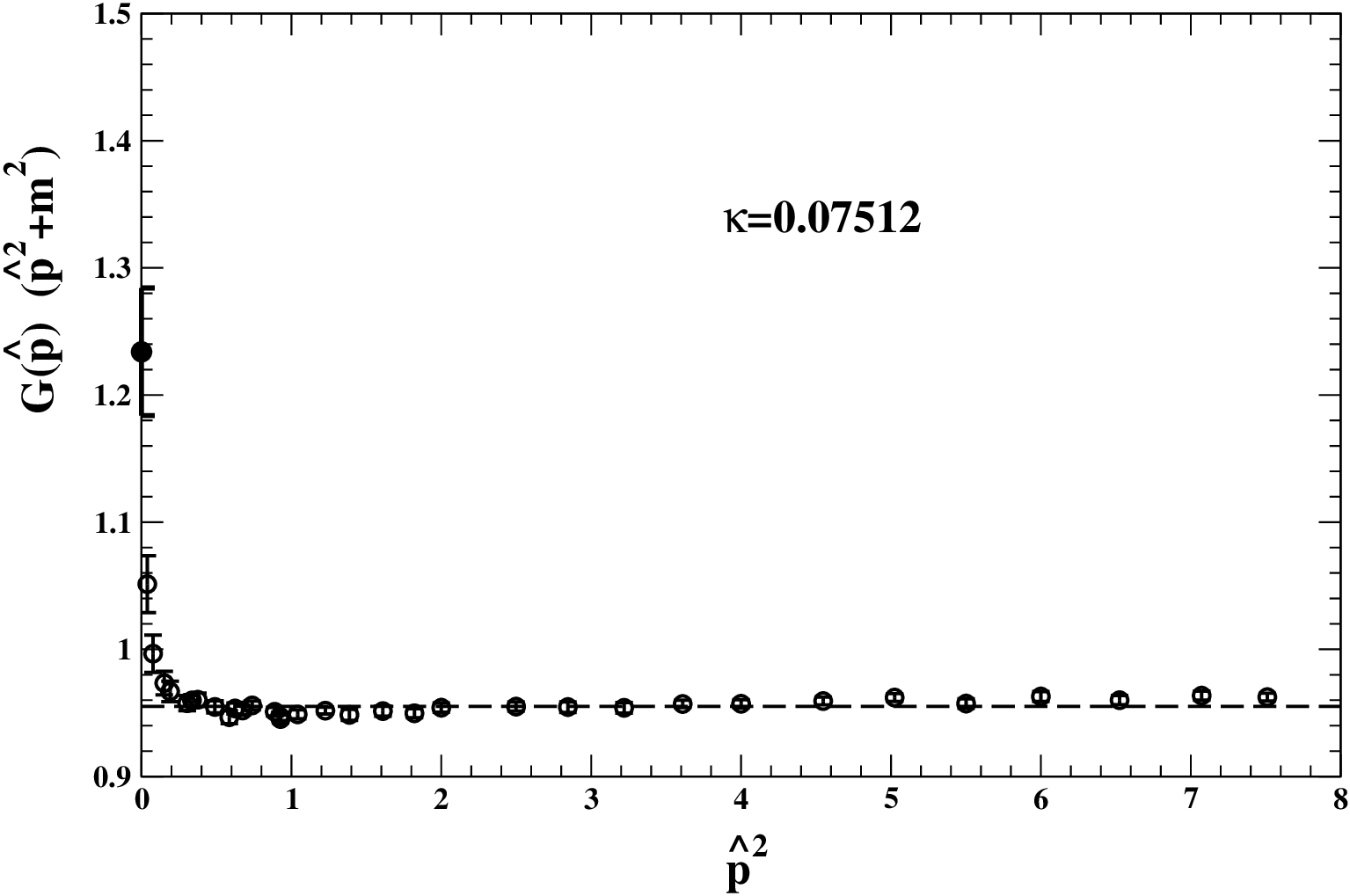}
\caption{\it The lattice data for the re-scaled propagator in the
broken phase at $\kappa=0.07512$ as a function of the square lattice
momentum ${\hat p}^2$ with $\hat{p}_\mu =2 \sin p_\mu/2$. The fitted
mass is $m_{\rm{latt}}=0.2062(41)$ and the dashed line indicates the
value of $Z_{\rm{prop}}=0.9551(21)$. The zero-momentum full point is
$Z_\varphi=(2\kappa\chi_{\rm{latt}})m^2_{\rm{latt}}=1.234(50)$. Data
are taken from Ref.~\cite{Cea:1999kn}. } \label{0.07512}
\end{figure}

\begin{figure}[ht]
\centering
\includegraphics[width=0.8\textwidth,clip]{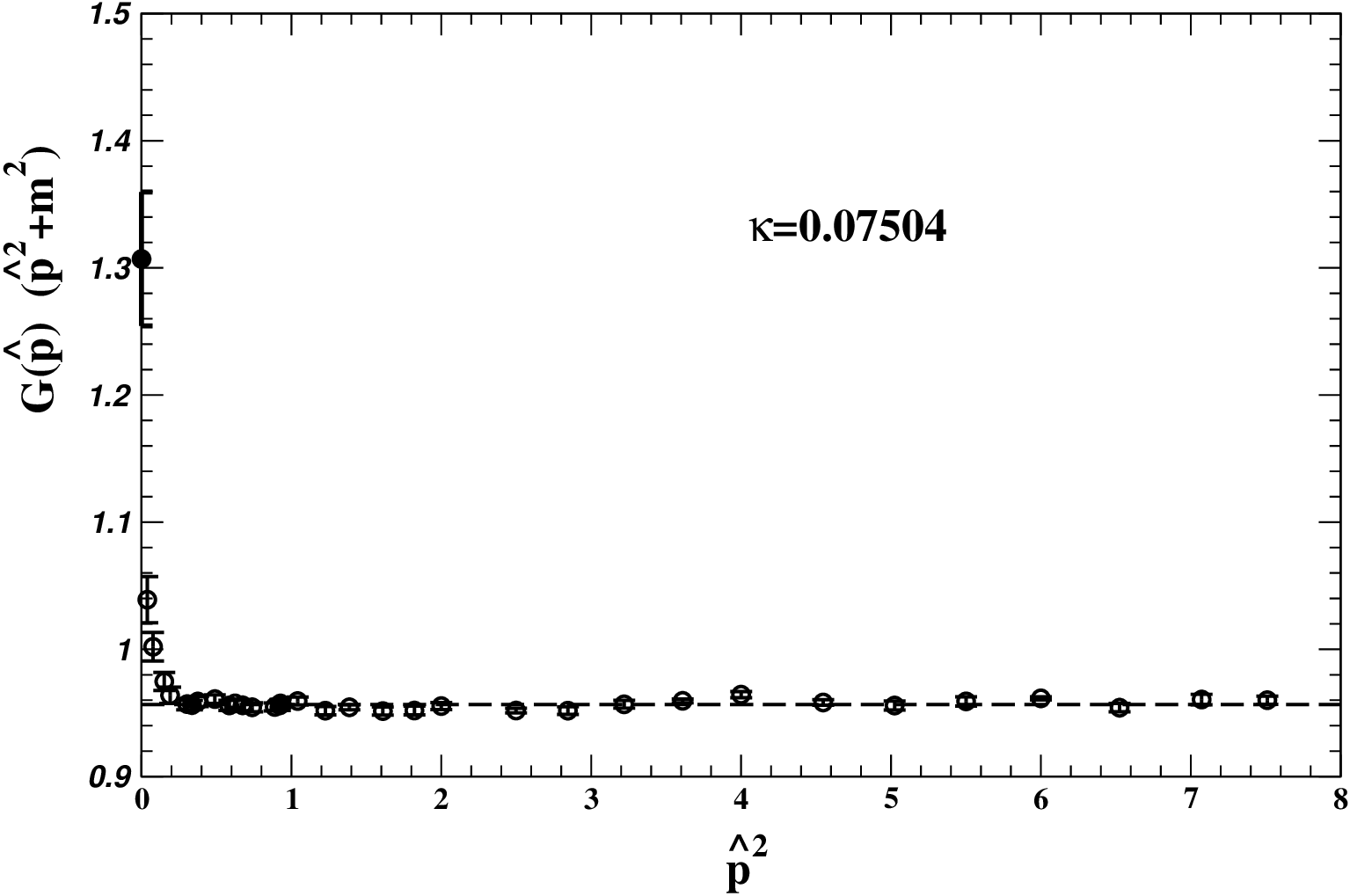}
\caption{\it The lattice data for the re-scaled propagator in the
broken phase at $\kappa=0.07504$ as a function of the square lattice
momentum ${\hat p}^2$ with $\hat{p}_\mu =2 \sin p_\mu/2$. The fitted
mass is $m_{\rm{latt}}=0.1723(34)$ and the dashed line indicates the
value of $Z_{\rm{prop}}=0.9566(13)$. The zero-momentum full point is
$Z_\varphi=(2\kappa\chi_{\rm{latt}})m^2_{\rm{latt}}=1.307(52)$. Data
are taken from Ref.~\cite{Cea:1999kn}. } \label{0.07504}
\end{figure}

This analysis was first performed in Ref.\cite{Cea:1999kn} for both
symmetric and broken phase. The data for the connected propagator
$2\kappa G(p)$ were first fitted to the 2-parameter form \BE
\label{twoparameter} G_{\rm{fit}}(p)=\frac{Z_{\rm{prop}} }{{\hat
p}^2 +m^2_{\rm{latt}} }\EE in terms of the squared lattice momentum
${\hat p}^2$ with $\hat{p}_\mu=2 \sin p_\mu/2$. The data were then
plotted after a re-scaling by the factor $({\hat
p}^2+m^2_{\rm{latt}})$. In this way, deviations from constancy
become clearly visible and indicate how well a given lattice mass
can describe the data down to $p\to 0$.

The results for the symmetric phase, in Fig.\ref{0.074} at
$\kappa=0.074$, show that, there, a single lattice mass works
remarkably well in the whole range of momentum down to $p=0$. Also
$Z_\varphi=(2\kappa)m^2_{\rm{latt}}\chi_{\rm{latt}}=0.9702(91)$
agrees very well with the fitted $Z_{\rm{prop}}=0.9682(23)$.

\begin{figure}[ht]
\centering
\includegraphics[width=0.7\textwidth,clip]{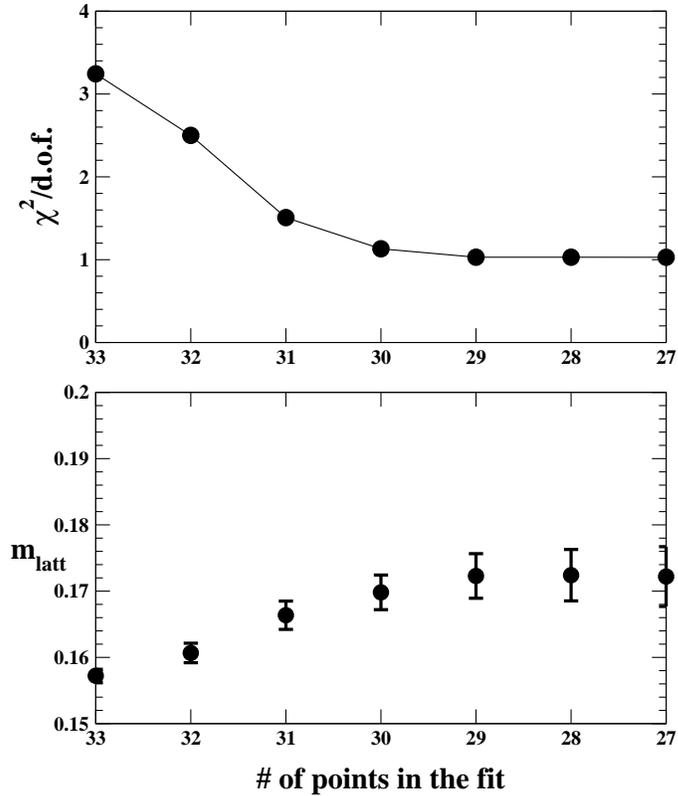}
\caption{\it For $\kappa=0.07504$ we show the value of the
normalized chi-square and the fitted lattice mass depending on the
number of points included in the high-energy region. Data are taken
from Ref.~\cite{Cea:1999kn}. } \label{chisquare}
\end{figure}

In Figs.\ref{0.07512} and \ref{0.07504} we then report the analogous
plots for the broken-symmetry phase at $\kappa=0.07512$ and
$\kappa=0.07504$ for $m_{\rm{latt}}=$ 0.2062(41) and
$m_{\rm{latt}}=$ 0.1723(34) respectively. As one can see, the fitted
lattice mass describe well the data for not too small values of the
momentum but in the $p\to 0$ limit the deviation from constancy
becomes highly significant statistically. To make this completely
evident, we show in Fig.\ref{chisquare} the normalized chi-square
vs. the number of points included in the fit.

Notice that the two quantities
$Z_\varphi=(2\kappa)m^2_{\rm{latt}}\chi_{\rm{latt}}=1.234(50)$ and
$Z_\varphi=(2\kappa)m^2_{\rm{latt}}\chi_{\rm{latt}}=1.307(52)$
respectively are now very different from the corresponding
quantities $Z_{\rm{prop}}=0.9551(21)$ and $Z_{\rm{prop}}=0.9566(13)$
obtained from the higher-momentum fits. Also, the value of
$Z_\varphi$ increases by approaching the critical point as expected.

The whole issue was thoroughly re-analyzed by Stevenson
\cite{Stevenson:2005yn} in 2005. For an additional check, he also
extracted propagator data from the time-slices for the connected
correlator measured by Balog et al. \cite{Balog:2004zd} for
$\kappa=0.0751$. He found that their higher-momentum data were
requiring a mass value $ m_{\rm{latt}}\sim 0.2$ but, again, see his
Fig.6(d), this mass could not describe the very low momentum points,
exactly as in our Figs.\ref{0.07512} and \ref{0.07504}. In
connection with the susceptibility $\chi_{\rm{latt}}=206.4(1.2)$
measured by Balog et al. at $\kappa=0.0751$ (see their Table 3),
this gives $Z_\varphi=(2\kappa\chi_{\rm{latt}})m^2_{\rm{latt}}\sim
1.24$ in very good agreement with our determination
$Z_\varphi=(2\kappa\chi_{\rm{latt}})m^2_{\rm{latt}}=1.234(50)$ at
the very close point $\kappa=0.07512$.

Therefore, data collected by other groups were confirming that in
the broken-symmetry phase $M_h\equiv m_{\rm{latt}}$, obtained from a
fit to the higher-momentum propagator data, and $m_h=(2\kappa
\chi_{\rm{latt}})^{-1/2}$ become more and more different in the
continuum limit.

\begin{table}[htb]
\tbl{The values of the susceptibility at various $\kappa$. The
results for $\kappa=0.07512$ and $\kappa=0.07504$ are taken from
ref.\cite{Cea:1999kn}. The result for $\kappa=0.0751$ is taken from
ref.\cite{Balog:2004zd} while the other value at $\kappa=0.0749$
derives from our new simulations on a $76^4$ lattice.}
{\begin{tabular}{@{}ccc@{}} \toprule
$\kappa$   &lattice &$\chi_{\rm{latt}}$  \\
\hline
0.07512     &$32^4$  &193.1(1.7)               \\
\hline
0.0751     &$48^4$  &206.4(1.2)               \\
\hline 0.07504    &$32^4$ &293.38(2.86)               \\
\hline 0.0749     &$76^4$ & 1129(24) \\
\botrule
\end{tabular}}
\end{table}

However since this is still not generally appreciated, and to
emphasize the phenomenological implications, we will now display
more precisely the predicted logarithmic increase of $Z_\varphi$. To
this end, we will show that the lattice data give consistent values
of the proportionality constant $c_2$ in Eq.(\ref{basicnew}) \BE
\label{cc2} Z_\varphi=\frac{M^2_h}{m^2_h} \equiv
(2\kappa\chi_{\rm{latt}})m^2_{\rm{latt}} \sim \frac{ L }{c_2} \EE
where $L\equiv \ln(\Lambda_s/m_{\rm{latt}})$. This requires to
compute the combination \BE \label{c2} m_{\rm{latt}} \sqrt{
\frac{2\kappa \chi_{\rm{latt}} }{\ln(\pi/am_{\rm{latt}}) }} \equiv
\frac{1}{\sqrt {c_2}} \EE where we have replaced the cutoff
$\Lambda_s \sim (\pi/a)$ in terms of the lattice spacing $a$. In
this derivation, no additional theoretical inputs (such as
definitions of renormalized mass and coupling constant) are needed.
The only two ingredients are i) the direct measurement of the
susceptibility and ii) the direct measurements of the connected
propagator. The higher-momentum region reproduced by the
two-parameter form Eq.(\ref{twoparameter}) is determined by the data
themselves and used to extract $m_{\rm{latt}}$.

\begin{table}[htb]
\tbl{The values of $m_{\rm{latt}}$, as obtained from a direct fit to
the higher-momentum propagator data, are reported together with the
other quantities entering the determination of the coefficient $c_2$
in Eq.(\ref{c2}). The entries at $\kappa=0.07512$ and
$\kappa=0.07504$ are taken from ref.\cite{Cea:1999kn}. The
susceptibility at $\kappa=0.0751$ is directly reported in
ref.\cite{Balog:2004zd}. The corresponding mass at $\kappa=0.0751$
was extracted by Stevenson \cite{Stevenson:2005yn} (see his
Fig.6(d)) by fitting to the higher-momentum data of
ref.\cite{Balog:2004zd}. The two entries at $\kappa=0.0749$, from
our new simulations on a $76^4$ lattice, refer to higher-momentum
fits for ${\hat p}^2>0.1$ and ${\hat p}^2>0.2$ respectively.}
{\begin{tabular}{ccccc}\toprule $\kappa$   &  $m_{\rm{latt}}$  & $
(2\kappa \chi_{\rm{latt}})^{1/2}$ &~~ $[\ln(\Lambda_s/m_{\rm{latt}}
)]^{-1/2}$ &
~~$(c_2)^{-1/2}$\\
\hline
0.07512 & 0.2062(41)    &  5.386(23)& 0.606(2) & 0.673(14)                \\
\hline 0.0751 & $\sim 0.200$     &  5.568(16) & $\sim 0.603$ & $\sim
0.671$\\
\hline 0.07504 & 0.1723(34)    &  6.636(32)& 0.587(2) & 0.671(14)                \\
\hline 0.0749 & 0.0933(28)    &  13.00(14) & 0.533(2) & 0.647(20)                \\
\hline 0.0749 & 0.100(6)    &  13.00(14) & 0.538(4) & 0.699(42)                \\
\botrule
\end{tabular}}
\end{table}

\begin{figure}[ht]
  \centering
  \includegraphics[width=0.8\textwidth,clip]{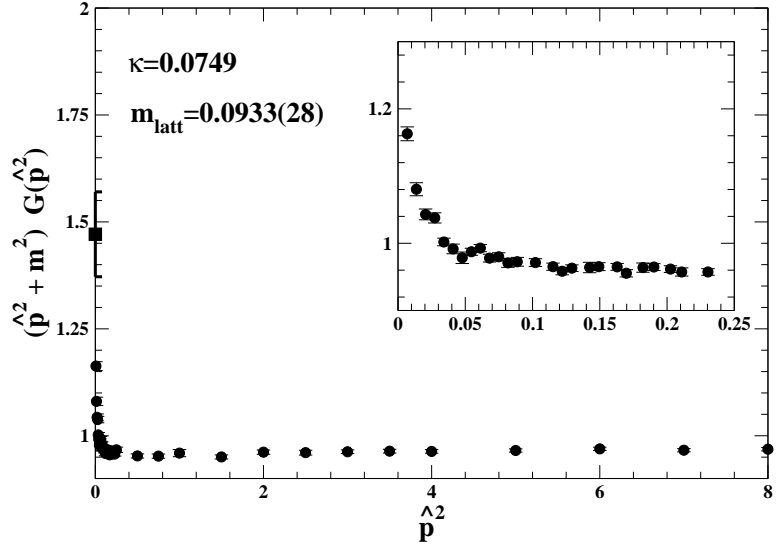}
  \caption{\it The propagator data, at
$\kappa=0.0749$, rescaled with the lattice mass
$m_{\rm{latt}}=0.0933(28)$ obtained from the fit to all data with
${\hat p}^2>0.1$. The square at $p=0$ is $Z_\varphi=
m^2_{\rm{latt}}(2\kappa\chi_{\rm{latt}})= 1.47(9)$.} \label{0933}
\end{figure}

\begin{figure}[ht]
  \centering
   \includegraphics[width=0.95\textwidth,clip]{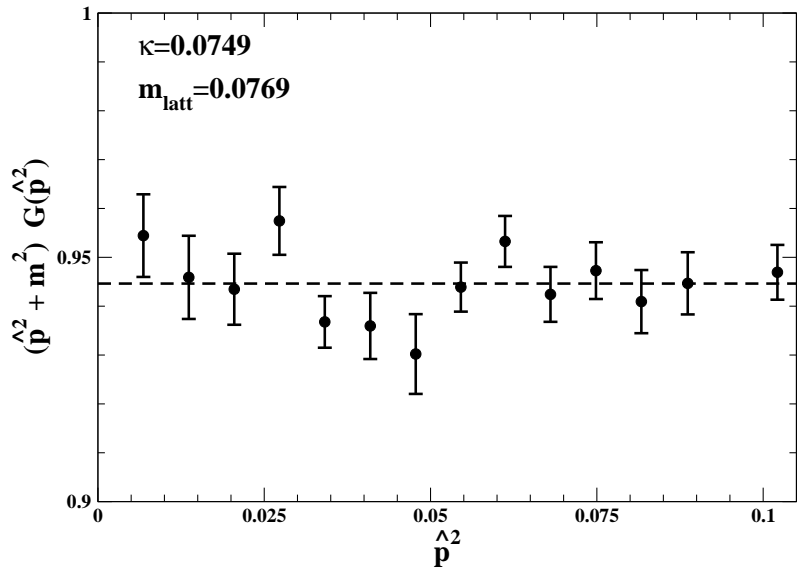}
   \caption{\it The propagator data at
$\kappa=0.0749$ for ${\hat p}^2<0.1$. The lattice mass used here for
the rescaling was fixed at the value
$m_h\equiv(2\kappa\chi_{\rm{latt}})^{-1/2}=0.0769$. } \label{susce}
\end{figure}

We give first in Table 1 the measured values of the lattice
susceptibility at various $\kappa$ (well within the scaling region).
We then report in Table 2 the fitted $m_{\rm{latt}}$ together with
the other quantities entering the determination of the coefficient
$c_2$ in Eq.(\ref{c2}). The spread of the central values at
$\kappa=0.0749$ reflects the theoretical uncertainty in the choice
of the higher-momentum range, ${\hat p}^2>0.1$ and ${\hat p}^2>0.2$
respectively. Only the region ${\hat p}^2<0.1$ cannot be
consistently considered with the rest of the data, see
Fig.\ref{0933}. In this low-momentum range the propagator data would
in fact require the same mass parameter
$m_h=(2\kappa\chi_{\rm{latt}})^{-1/2}=0.0769$ fixed by the inverse
susceptibility, see Fig.\ref{susce}.

The reason of this uncertainty is that, differently from the
simulations at $\kappa=0.07512$ and $\kappa=0.07504$, this
higher-momentum range cannot be uniquely determined by simply
imposing a normalized chi-square of order unity as in
Fig.\ref{chisquare}. To this end, in fact, statistical errors should
be reduced by, at least, a factor of $2$ with a corresponding
increase of the CPU time by a factor $4$. Due to the large size
$76^4$ of the lattice needed to run a simulation at $\kappa=0.0749$,
this increase in statistics would take several additional months.

Nevertheless, with our present statistics this type of uncertainty
can be translated into the average estimate $m_{\rm{latt}} \sim
0.096 (3)$ at $\kappa=0.0749$, or $1/\sqrt{c_2}\sim 0.67 \pm 0.02$.
In turn, besides the statistical errors, this is equivalent to a
systematic error $\pm 0.02 $ in the final average  \BE
\label{c2final} \frac{1}{\sqrt {c_2}}=0.67 \pm 0.01({\rm stat}) \pm
0.02({\rm sys}) \EE With this determination, we can then compare
with the L\"uscher-Weisz scheme \cite{Luscher:1987ek} where mass $m_{\rm R}$,
coupling constant \footnote{In the L\"uscher-Weisz paper the scalar
self coupling is called $g$. However, here, to avoid possible
confusion with the gauge couplings we will maintain the traditional
notation $\lambda$.} $\lambda_{\rm R}$ and weak scale $\langle
\Phi\rangle$ are related through the relation \BE \label{mlw}
\frac{m^2_{\rm R} }{\langle \Phi\rangle^2}= \frac{\lambda_{\rm R}
}{3} \EE and the mass is expressed in terms of the zero-momentum
propagator as \BE \frac{Z_{\rm R} }{m^2_{\rm R}} = G(p=0) = 2\kappa
\chi= \frac{1 }{m^2_h} \EE through a perturbative rescaling $Z_{\rm
R} \lesssim1$.

Traditionally, Eq.(\ref{mlw}) has been used to place upper bounds on
the Higgs boson mass depending on the value of $\lambda_{\rm R}\sim
(1/L)$ and thus on the magnitude of $\Lambda_s$. Instead, in our
case, where $M^2_h \sim L m^2_h \sim L m^2_{\rm R} $, it can be used
to express the value of $M_h$ in units of $\langle \Phi\rangle$
because the two quantities now scale uniformly, see
Eq.(\ref{kfinite}). Since our estimate of the $M_h-m_h$ relation
just takes into account the leading-order logarithmic effect, in a
first approach, we will neglect the non-leading quantity $Z_{\rm R}$
and, as sketched at the end of Sect.2, approximate $m_{\rm R}\sim
m_h$. Therefore, by using the leading-order relation $(m_h/\langle
\Phi\rangle)^2\sim 16 \pi^2/(9L)$, Eq.(\ref{cc2}) and the average
value Eq.(\ref{c2final}), the logarithmic divergent $L$ drops out
and we find \BE \frac{M_h}{\langle \Phi\rangle}=
\sqrt{\frac{m^2_h}{\langle \Phi\rangle^2 } \frac{M^2_h}{m^2_h }}\sim
\sqrt{ \frac{16\pi^2}{9L } \frac{L}{c_2 } }= 2.81 \pm 0.04({\rm
stat}) \pm 0.08({\rm sys})  \EE or, for $ \langle \Phi\rangle \sim$
246 GeV, \BE \label{leading} M_h = 690 \pm 10 ({\rm stat}) \pm
20({\rm sys}) ~~\rm{GeV} \EE We observe that the above value is
slightly smaller but consistent with our previous estimate
\cite{Cea:2003gp,Cea:2009zc} \BE\label{old} M_h = 754 \pm 20 ({\rm
stat}) \pm 20({\rm sys}) ~~\rm{GeV} \EE This had been obtained,
within the same L\"uscher-Weisz scheme, but using instead the full
chain \BE \label{modified} \frac{M_h}{\langle \Phi\rangle} =
\sqrt{\frac{M^2_h }{m^2_h} \frac{m^2_h }{m^2_{\rm R}} \frac{m^2_{\rm
R} }{ \langle \Phi\rangle^2 }}=\sqrt{ \frac{Z_\varphi}{Z_{\rm R}}
\frac{\lambda_{\rm R} }{3}} \EE and thus account for both the
logarithmic divergent $Z_\varphi $ and the non-leading correction
$Z_{\rm R}$.

This old estimate Eq.(\ref{old}) can now be compared with our new
determination of $Z_\varphi$ from the direct measurement of the
lattice propagator. To eliminate any explicit dependence on the
lattice mass it is convenient to introduce the traditional divergent
log used to describe the continuum limit of the Ising model
\cite{ZinnJustin:2002ru} \BE \label{lkappa}L(k)= \frac{1}{2}\ln
\frac{\kappa_c}{\kappa -\kappa_c} \EE and define a set of values \BE
\label{zkappa} Z_\varphi\equiv \frac{ L(k)}{c_2} \EE at the various
$\kappa$. By using our Eq.(\ref{c2final}) and
$\kappa_c=0.0748474(3)$, all entries needed in Eq.(\ref{modified})
are reported in Table 3. Then, by averaging at the various $\kappa$,
the new determination $M_h\sim 752 \pm 20$ GeV is the same value
Eq.(\ref{old}) obtained in refs.\cite{Cea:2003gp,Cea:2009zc}.

\begin{table}[htb]
\tbl{We report the original L\"uscher-Weisz entries
\cite{Luscher:1987ek} $\lambda_{\rm R}$  and  $Z_{\rm R}$,  the
rescaling $\sqrt{Z_\varphi} \equiv \sqrt{\frac{L(k)}{c_2}}$, with
$L(k)$ as in Eq.(\ref{lkappa}) and $1/\sqrt{c_2}=0.670 \pm 0.023$ as
in Eq.(\ref{c2final}), together with the resulting $M_h$ from
Eq.(\ref{modified}). } {\begin{tabular}{ccccc}\toprule $\kappa$  &
$\lambda_{\rm R}$  & $Z_{\rm R}$ &
$\sqrt{Z_\varphi}$ & $M_h$ ({\rm GeV})\\
\hline
0.0759     &  27(2)  &  0.929(14) &0.98(3)    & 751 (37)       \\
\hline
0.0754      &24(2)  &  0.932(14) &1.05(4)  & 757 (40)            \\
\hline 0.0751   & 20(1)  &  0.938(12)& 1.13(4)  & 742 (33)               \\
\hline 0.0749    & 16.4(9)  & 0.944(11) &1.28(5) & 758 (34)  \\
\botrule
\end{tabular}}
\end{table}

One may object that the new precise $\kappa_c$ is marginally
consistent with the old value $0.07475(7)$ used originally by
L\"uscher-Weisz \cite{Luscher:1987ek} to compute the $\lambda_{\rm R}$'s and
$Z_{\rm R}$'s reported in Table 3. However, $Z_{\rm R}$ is a very
slowly varying, non-leading quantity whose dependence on the
critical point is well within the uncertainties reported in Table 3.
Also, the dependence of $\lambda_{\rm R}$ on the various mass scales
is only logarithmic and possible differences are further flattened
because only $\sqrt{\lambda_{\rm R}}$ enters the determination of
$M_h$ \footnote{With a critical $\kappa_c=0.074848$ very close to
the present most precise determination $\kappa_c=0.0748474(3)$, the
$\lambda_{\rm R}$'s were re-computed by Stevenson
\cite{Stevenson:2005yn}, see his Fig.1 (f). His new central values are
about $\lambda_{\rm R}=$ 30, 25, 21, 16.7 for $\kappa=$ 0.0759,
0.0754, 0.0751 and 0.0749 respectively and thus within the
uncertainties reported in Table 3. In any case, the average $+2.7\%$
increase in the central value of $M_h$ remains within the $\pm 20$
GeV systematic error reported in Eq.(\ref{old}).}.

We thus conclude that, either with the original estimate of
refs.\cite{Cea:2003gp,Cea:2009zc} or with our new determination of
$Z_\varphi$ in Table 3, Eq.(\ref{modified}) remains as an
alternative approach to $M_h$ which has its own motivations and
takes also into account the average $+3\% $ effect embodied in
$\sqrt{Z_{\rm R}}\sim 0.97$. In this perspective,
Eqs.(\ref{leading}) and (\ref{old}) could be combined in a final
estimate \BE \label{final} M_h= 720 \pm 30~{\rm GeV} \EE which
incorporates the various statistical and theoretical uncertainties.

\section{Summary and outlook}

In the first version of the theory, with a classical scalar
potential, the sector inducing SSB was quite distinct from the
remaining self-interactions of the Higgs field induced through its
gauge and Yukawa couplings. In this paper, we have adopted a similar
perspective but, following most recent lattice simulations,
described SSB in $\lambda\Phi^4$ theory as a weak first-order phase
transition.

In the approximation schemes we have considered, there are two
different mass scales. On the one hand, a mass $m_h$ defined by the
quadratic shape of the effective potential at its minimum and
related to the zero-momentum self-energy $\Pi(p=0)$. On the other
hand, a second mass $M_h$, defined by the zero-point energy which is
relevant for vacuum stability and related to a typical average value
$ \langle \Pi(p)\rangle $ at larger $|p|$.

So far, these two scales have always been considered as a single
mass but our results indicate instead the order of magnitude
relation $M^2_h\sim m^2_h L\gg m^2_h$, where $L=\ln (\Lambda_s/M_h)$
and $\Lambda_s$ is the ultraviolet cutoff of the scalar sector which
induces SSB. We have checked this two-scale structure with lattice
simulations of the propagator and of the susceptibility in the 4D
Ising limit of the theory. These confirm that, by approaching the
critical point, $M^2_h$, as extracted from a fit to the
higher-momentum propagator data, increases logarithmically in units
of $m^2_h$, as defined from the inverse zero-momentum susceptibility
$|\Pi(p=0)|=(2\kappa \chi)^{-1}$. At the same time, see
Fig.\ref{susce}, $m_h=(2\kappa \chi)^{-1/2}$ is the right mass to
describe the propagator in the low-momentum region. Therefore, in a
cutoff theory where both $m_h$ and $M_h$ are finite, one should
think of the scalar propagator as a smooth interpolation between
these two masses.

With the aim of extending our description of SSB to more ambitious
frameworks, we have also developed in Sect.2 a RG-analysis which, in
principle, could also be extended to the $\Lambda_s \to \infty$
limit and introduces two invariants ${\cal I }_1$ and ${\cal I }_2$.
The former is related to the vacuum energy ${\cal E }\sim -M^4_h$,
through the relation ${\cal I }_1=M_h$. The latter is the natural
candidate to represent the weak scale $\langle \Phi \rangle\sim$ 246
GeV through the relation ${\cal I }_2= \langle \Phi \rangle$.

Therefore since, differently from $m_h$, the larger mass $M_h$
remains finite in units of $\langle \Phi \rangle$ in the continuum
limit, one can write a proportionality relation, say $M_h=K \langle
\Phi \rangle$, and extract the constant $K$ from lattice
simulations. As discussed in Sect.5, this leads to our final
estimate $M_h \sim 720 \pm 30 $ GeV which incorporates various
statistical and theoretical uncertainties.

The existence of two masses in our picture of SSB leads to exploit
the natural identification of our lower mass $m_h$ with the present
experimental value 125 GeV. In this case, we obtain \BE \frac{
M_h}{125~ {\rm GeV} } \sim \sqrt{ \frac{ L}{c_2} }\EE so that from
\BE M_h \sim \frac{ 4\pi\langle\Phi\rangle }{3 \sqrt{c_2} } \EE we
find $\sqrt{ L} \sim 8.25 $. When taken at face value, this would
imply a scalar cutoff $\Lambda_s\sim 2.6 \cdot 10 ^{32}$ GeV which
is much larger than the Planck scale. But, as pointed out in the
footnote before Eq.(\ref{I1}), this may be just a cutoff artifact
because to obtain the same physical $M_h$ from Eq.(\ref{basic}) and
Eq.(\ref{I1}) one should use vastly different values of the
ultraviolet cutoff.

Instead, as emphasized in the Introduction, our aim was to give a
cutoff-independent description of symmetry breaking in
$\lambda\Phi^4$ theory, i.e. a description that could also remain
valid in the $\Lambda_s \to \infty$ limit.  In this perspective, for
an experimental check of our picture, we should first look at the
{\it cutoff-independent} $M_h-\langle \Phi \rangle$ relation. Since
this would imply the existence of a new scalar resonance around 700
GeV, we will now briefly recall some experimental signals from LHC
that may support this prediction. The $M_h-m_h$ relative magnitude
will be re-discussed afterwards by making use of a physical,
measurable quantity.

Let us start with the 2-photon channel. At the time of 2016, both
ATLAS \cite{Aaboud:2016tru} and CMS \cite{Khachatryan:2016hje} experiments
reported an excess of events in the 2-photon channel that could
indicate a new narrow resonance around 750 GeV. The collisions were
recorded at center of mass energy of 8 and 13 TeV and the local
statistical significance of the signal was estimated to be 3.8 sigma
by ATLAS and 3.4 sigma by CMS. Later on, with more statistics, the
two Collaborations reported a considerable reduction in the observed
effect. For ATLAS \cite{Aaboud:2017yyg} the local deviation from the
background-only hypothesis was reduced to 1.8 sigma while for CMS
\cite{Khachatryan:2016yec}, the original 3.4 sigma effect was now lowered
to about 1.9 sigma. Yet, in spite of the reported modest statistical
significance, if one looks at the 2-photon invariant mass
distribution in figure 2a of ATLAS \cite{Aaboud:2017yyg}, an excess of
events at about 730 GeV is clearly visible. Interestingly, this
excess is immediately followed by a strong decrease in the number of
events. This may indicate the characteristic $(M^2-s)$ effect due to
the (negative) interference of a resonance of mass $M$ with a
non-resonating background. These last papers were published in 2017
and the total integrated luminosity was 36 fb$^{-1}$ (12.9 + 19.7 +
3.3) for CMS and 36.7 fb$^{-1}$ for ATLAS. This is just a small
fraction of the full present statistics of about 140 fb$^{-1}$ per
experiment.

Let us now consider the ``golden'' 4-lepton channel at large values
of the invariant mass $m_{4l}>600$ GeV. For the latest paper by
ATLAS \cite{Aaboud:2017rel}, with a statistics of 36.1 fb$^{-1}$, one
can look at their figure 4a. Again, as in their corresponding
2-photon channel (the mentioned figure 2a of \cite{Aaboud:2017yyg}),
there is a clean excess of events for $m_{4l}=$ 700 GeV where the
signal exceeds the background by about a factor of three. At the
closest points, 680 and 720 GeV, the signal becomes consistent with
the background within 1 sigma but the central values are still
larger than the background by a factor of two. The other paper by
CMS \cite{CMS:2018mmw} refers to a statistics of 77.4 fb$^{-1}$ but the
results in the region $m_{4l}\sim$ 700 GeV, illustrated in their
Fig.9, cannot be easily interpreted.

However, here, an independent analysis of these data by Cea
\cite{Cea:2018tmm} can greatly help. The extraction of the CMS data
and their combination with the ATLAS data presented in Figures 1 and
2  of ref. \cite{Cea:2018tmm} indicates an evident excess in the
4-lepton final state with a statistical significance of about 5
sigma. The natural interpretation of this excess would be in terms
of a scalar resonance, with a  mass of about $700$ GeV, which decays
into two $Z$ bosons and then into leptons. We emphasize that one
does not need to agree with Cea's theoretical model to appreciate
his analysis of the data. Therefore, if this excess will be
confirmed, it could represent the second heavier mass scale
discussed in our paper. We emphasize that the statistical sample
used in \cite{Cea:2018tmm} is the whole official set of data
available at present, namely 113.5 fb$^{-1}$ (36.1 for ATLAS + 77.4
for CMS). Again, as for the 2-photon case, this is still far from
the nominal collected luminosity of about 140 fb$^{-1}$ per
experiment.

In this situation, where only a small fraction of the full
statistics has been made available, further speculations on the
characteristics of a hypothetical heavy mass state at 700 GeV may be
premature. Nevertheless, even though this scale is not far from the
usual triviality bounds, the actual situation we expect is very
different. In fact these bounds have been obtained for $M_h\lesssim
\Lambda_s$ while we are now considering a corner of the parameter
space, i.e. large $M_h$ with $M_h \ll \Lambda_s$, that does not
exist in the conventional treatment. For this reason the
phenomenology of such heavy resonance (i.e. its production cross
sections and decay rates) may differ sizeably from the perturbative
expectations. In particular, differently from the low-mass state at
125 GeV, the decay width of the heavy state into longitudinal vector
bosons will be crucial to determine the strength of the scalar
self-interaction. We thus return to the previous issue concerning
the relative magnitude of $M_h$ and $m_h$.

From the experimental ATLAS + CMS papers that we have considered,
the total width of this hypothetical heavy resonance can hardly
exceed 40 GeV. For a mass of 720 GeV, about 30 GeV of this width,
those into heavy and light fermions, gluons, photons...would
certainly be there. Thus, the decay width into W's and Z's should be
of the order of 10 GeV, or less. The observation of such a heavy but
narrow resonance would then confirm the scenario of
ref.~\cite{Castorina:2007ng} where, with a heavy Higgs particle,
re-scattering of longitudinal vector bosons was effectively reducing
their large tree-level coupling and thus the decay width in that
channel. In the language of the present paper, this could be
expressed by saying that the tree-level estimate $\Gamma_0(h \to V_L
V_L) \sim M^3_h G_{\rm Fermi}\sim$ 175 GeV becomes the much smaller
value $\Gamma(h \to V_L V_L) \sim M_h (m^2_h G_{\rm Fermi})$ where
$M_h$ is from phase space and $m^2_h G_F$ is the reduced strength of
the interaction. If $M_h$ is close to 720 GeV and the mass $m_h$
needed for the reduction of the width is close to 125 GeV, say a
width into vector bosons of the order of 5 GeV, this would then
close the circle and lead to the identification $m_h \sim 125$ GeV.

Finally, the simultaneous presence of two different mass scales in
the Higgs field propagator would also require some interpolating
form, of the type Eq.(\ref{interpolation}), in the loop corrections.
Since some precision measurements (e.g. the b-quark forward-backward
asymmetry or the value of $\sin^2\theta_w$ from neutral current
experiments \footnote{For a general discussion of the various
quantities and of systematic errors see ref.\cite{Chanowitz:2009dz}.}) still
point to a rather large Higgs particle mass, with about 3-sigma
discrepancies, this could provide an alternative way to improve the
overall quality of a Standard Model fit.

%\bibliography{higgs}

\end{document}